# Cascaded emission of single photons from the biexciton in monolayered WSe$_2$

Yu-Ming He[1,4], Oliver Iff[1], Nils Lundt[1], Vasilij Baumann[1], Marcelo Davanco[2], Kartik Srinivasan[2], Sven Höfling[1,3] and Christian Schneider[1,*]

[1]*Technische Physik and Wilhelm-Conrad-Röntgen-Research Center for Complex Material Systems, Universität Würzburg, D-97074 Würzburg, Am Hubland, Germany.*

[2]*Center for Nanoscale Science and Technology, National Institute of Standards and Technology, Gaithersburg, MD 20899, USA*

[3]*SUPA, School of Physics and Astronomy, University of St. Andrews, St. Andrews KY 16 9SS, United Kingdom*

[4]*Hefei National Laboratory for Physical Sciences at the Microscale and Department of Modern Physics, & CAS Center for Excellence and Synergetic Innovation Center in Quantum Information and Quantum Physics, University of Science and Technology of China, Hefei, Anhui 230026, China*

*Correspondence and requests for materials should be addressed to Christian Schneider (christian.schneider@physik.uni-wuerzburg.de)

**Monolayers of transition metal dichalcogenide materials emerged as a new material class to study excitonic effects in solid state, since they benefit from enormous coulomb correlations between electrons and holes. Especially in WSe$_2$, sharp emission features have been observed at cryogenic temperatures, which act as single photon sources. Tight exciton localization has been assumed to induce an anharmonic excitation spectrum, however, the evidence of the hypothesis, namely the demonstration of a localized biexciton, is elusive. Here, we unambiguously demonstrate the existence of a localized biexciton in a monolayer of WSe$_2$, which triggers an emission cascade of single photons. The biexciton is identified by its time-resolved photoluminescence, superlinearity and distinct polarization in micro-photoluminescence experiments. We evidence the cascaded nature of the emission process in a cross-correlation experiment, which yields a strong bunching**

behavior. Our work paves the way to a new generation of quantum optics experiments with two-dimensional semiconductors.

**Introduction**

Increasing interest and need for secure communications[1], precision measurements, metrology[2] and optical quantum emulation[3] explain the ever growing demand for non-classical sources of light. Ultra-compact implementations of such sources in solid state are particularly interesting, as they promise long-term stability, and outline possible ways for scalable integration, even in complex on chip quantum networks[4]. Regarding compactness, the ultimate limit is represented by a zero-dimensional emission center of non-classical light in atomic monolayers[5]. Transition metal dichalcogenides semiconductors have emerged as a new platform to study excitonic effects in two dimensions[6,7,8,9,10], in particular since they benefit from enormous coulomb correlations between electrons and holes as a result of reduced dielectric screening[11,12,13] and feature unique spinor properties[14]. Strong progress has been made in studying excitonic effects in two dimensional materials, including the exploration of the valley pseudospin dynamics[14,15], valley polarization control[16,17,18,19,20] and strong coupling effects[21,22]. While single photon emission from localized states in monolayers of transition metal dichalcogenides has recently been observed[23,24,25,26,27], the characteristic feature which determines an anharmonic excitation spectrum in an excitonic system, which is a well isolated biexcitonic state, has not been observed. In addition, strong spectral wandering and even blinking, induced by a supposedly noisy environment, made it debatable whether the full potential of such ultra-compact solid state single photon sources or pair sources based on mono-atomic layers can be harnessed. Recently, delocalized biexcitonic states in $WSe_2$ have been attributed to a superlinearly increasing, broad emission feature on the low energy side of the characteristic excitonic resonance[28]. However, these indications clearly outline the need for further unambiguous evidence of multiexcitonic complexes in monolayer materials.

Here, we demonstrate the existence of a localized biexciton in a monolayer of $WSe_2$ at cryogenic temperatures. The optical properties of the localized excitons are strongly improved by transferring them onto epitaxially grown semiconducting material. We routinely observe sharp emission features with linewidth < 70 µeV by means of non-resonant photoluminescence spectroscopy. The localized exciton-biexciton pair is identified by its characteristic polarization, the characteristic power dependency, and most importantly, the emission cascade is observed in single photon correlation studies.

**Results**

**Device description and optical characterization**

Figure 1a depicts a graphic illustration of our investigated device: It consists of a 250 nm thick GaInP layer, which has been grown on a semi-insulating GaAs substrate via Gas-Source molecular beam epitaxy. A single layer of $WSe_2$, mechanically exfoliated via commercial adhesive tape (Tesa brand) from a bulk crystal was transferred onto the atomically smooth and chemically inert GaInP layer with a polymer stamp. We did not apply any capping technique to the monolayer, to take advantage of the light extraction from the surface allocated emitters. Figure 1b illustrates a two-dimensional scanning image of the photoluminescence emission in the energy range between 1.525eV and 1.734eV. The bulk shape could be reproduced by the suppressed photoluminescence from the substrate in the energy range 1.525-1.55eV and the localized shining spots show the sharp peaks emission, which are randomly found at the edge of the flake.

A series of non-resonant micro photoluminescence spectra recorded at varying pump powers is shown in Fig. 1c). The single quantum emitter, which was excited by a 532 nm continuous wave laser at a sample temperature of nominally 4.5K, exhibits two pronounced emission features with a distinctly different power dependency. We note, that the two emission features, which have an energetic separation of 4.6 meV, have a spectral linewidth which is limited by the resolution of our spectrometer (70 µeV) at lowest pump powers. This outlines the excellent optical properties of our localized emitters, which are not affected by obvious effects of long term spectral wandering, as commonly observed for $WSe_2$ emitters on insulating substrates[23,24,25,26,27]. More details concerning the optical quality of our quantum emitters can be found in Supplementary Figure 1 and Supplementary Note 1. As we increase the pump power, the linewidth of our emission features steadily increases, which we attribute to power induced dephasing[29] (see Fig 1d). More importantly, as shown in Fig. 1e), we observe a distinctively different behavior of the integrated intensity of the two features, as we increase the excitation power. The intensity of the high energy signal (P2) increases less rapidly than the low energy signal (P1), and a characteristic saturation behavior of P1 is present as P2 still increases in intensity. We fit our data with a power law $I \sim P^x$, and find a powerlaw coefficient of $x = 0.84$ for P2 and a superlinear coefficient of $x = 1.42$ for P1. This pronounced sub- and superlinear behavior of the two emission features already outlines a possible emission cascade which is initialized by a biexcitonic state. With increasing sample temperature, P1 and P2 experience a characteristic broadening of the emission lines as well as a quenching of their intensity which is detailed in the Supplementary Figure 2 and Supplementary Note 2. A second pair of emission lines, which we found in our monolayer and which behaves qualitatively similar, is depicted in Supplementary Figure 3.

The exciton and biexciton could be further identified by the time-resolved photoluminescence of the localized emitter excited with a 3ps pulsed laser at 475nm (See Fig. 1f, and Supplementary Figure 4). Both P1 and P2 were fitted with a single exponential decay function with the time constant of $\tau = 0.793 \pm 0.017$ns for P1 and $\tau = 1.504 \pm 0.028$ns for P2. For the exciton and biexciton in semiconductor quantum dots, the time evolution of the probabilities for exciton (X) and biexciton (XX) could be modeled with the radiative transition functions[30] and the ratio $\tau_x / \tau_{xx}$ would be dependent on their electron-hole spatial wave function separation. A ratio $\tau_x / \tau_{xx} = 2$ is expected for excitons in the presence of a confining potential, and if the biexciton size was much larger than the exciton size. Here the extracted ratio ~ 1.897 corroborates our assignment of the exciton and biexciton lines. We point out, that in the monolayer WSe$_2$, the free biexciton diameter was expected to be 4 times larger than the free exciton size, which would translate into a similar ratio.

Resulting from the inherent selection rules of tightly localized excitons, emission from the biexcitonic cascade features a very characteristic polarization behavior. In fact, if the exciton is subject to a fine structure splitting, induced by structural anisotropies, the biexcitonic emission feature is typically subject to the same splitting, yet with opposite sign. This is a natural consequence of the emission cascade, which is sketched in Fig. 2a). We test the polarization features of our emission pair at low excitation powers. Indeed, for low pump powers, our two primary emission features of interest each split into doublets, which already indicates the presence of a fine structure in our system (Fig. 2b). We carry out polarization resolved spectroscopy by inserting a linear polarizer and a $\lambda/2$ wave-plate in our beam path, and study the intensity of each of these split peaks as a function of the polarizer orientation. Figure 2c depicts the intensity of each signal, which all feature a strong sinusoidal behavior suggesting a close-to perfect linear polarization of both the P1 and the P2 signal. This becomes even clearer as we normalize each peak, and plot the corresponding spectra in the contour graph in Fig. 2d. As expected from the emission cascade, the two possible branches of the cascade are separated by the characteristic fine structure splitting both in the X and the XX feature with an oscillation period shifted by a phase of $\pi$.

**Single photon correlation spectroscopy**

The key hypothesis, namely the presence of an emission cascade from the biexcitonic state to the crystal ground state, can be verified via photon correlation experiments. We prove the capability of our system to emit quantum light by measuring the second order autocorrelation function of the emission feature P2 under pulse-laser excitation, which we have attributed to the excitonic state. The corresponding autocorrelation histogram from the emission signal is shown in Fig 3a. The emission is spectrally filtered by a pair of band filters and then coupled into a fiber-based Hanbury-Brown and Twiss setup. Figure 3b shows the autocorrelation of the emission P1, which was attributed to the biexcitonic state and filtered out by a monochromator. The results around $\tau=0$ reveal $g^2(0)=0.397\pm0.04$ for P1 (XX) and $g^2(0)=0.218\pm0.06$ for P2 (X), which reach down well below 0.5 and therefore prove single photon emission from both states. Corresponding measurements under CW non-resonant excitation are shown in Supplementary Figure 5. Next, we measure the cross correlation between the emission signals P1 and P2. Figure 3c shows the photon-correlation measurements under the same pulse excitation between two peaks, from which we obtained the second-order correlation at zero delay of $g^2(0)=1.404\pm0.038$ (see Methods for details on the data analysis). The right inset of Fig. 3c shows the cross correlation between P1 and P2 under Continuous-Wave (CW) excitation. For $\tau<0$, we observe a characteristic antibunching, which transits to a value significantly above unity at small positive delays. This bunching effect definitely proves the cascaded nature of the single photon emission[34,35], and thus strongly suggests the presence of an emission cascade from a biexciton in our monolayer system.

**Discussion**

The microscopic properties of the localized biexciton and its selection rules in monolayers of $WSe_2$ require more work for clarification. The free biexciton states in $WSe_2$ monolayers have been predicted by variational calculations and experimental indications were previously discussed. Similarly to the expectations for free biexcitonic states in $WSe_2$, our lifetime measurements indicate that the localized biexcitons are probably larger than the exciton. Furthermore, a previous study of localized excitons in $WSe_2$ indicated the presence of a dark localized exciton as well as a free dark exciton state several meV below the bright exciton[37,38]. In our current study, we did not observe any significant contribution of a dark state, which would necessarily lead to a blinking in the defect emission, and thus to a long term bunching in the autocorrelation of the exciton emission. However, it will be interesting to see, if e.g. a temperature induced population of the dark exciton will lead to a modification of the observed quantum statistics in our cascaded system.

In conclusion, we have observed clear evidence for the presence of a localized biexcitonic complex in a monolayer of $WSe_2$. The successful observation of the localized biexciton is a consequence of the very clean emission spectra, which in turn are a result of our hybrid semiconductor-monolayer heterostructure . The emission signal features a superlinear increase of its intensity with excitation power, its fine structure splitting is of the same magnitude as the corresponding exciton, yet of opposite linear polarization, and most importantly, the characteristic emission cascade is unambiguously verified in a cross correlation experiment. Time resolved spectroscopy on the exciton-biexciton pair suggests that the biexcitonic Bohr radius is approximately four times the size of the exciton's spatial extension. We believe that this demonstration of a multi-particle complex in a monolayer material will give rise to a new plethora of interesting effects, such as the demonstration of valley entanglement. Furthermore, the high optical quality of our emitter on the surface of a sample it is a promising step towards the implementation of ultra compact, bright sources of single, indistinguishable[31] photons and even pairs of polarization and time-bin entangled photons[32]. Since the emitting dipole is allocated on a surface, rather than embedded in a high index medium, we believe that simple photonic architectures can be developed to obtain broadband quantum light sources with very large photon extraction[36].

**Optical measurements and data analysis**

The diagram of the experimental setup is shown in Supplementary Figure 6 and described in great detail in Supplementary Note 3. The sample was optically excited non-resonantly using a CW 532nm laser and a mode locked, up-converted Ti:Sapphire (475nm) with a pulse frequency of 82 MHz and a pulse length of 3 ps. A short pass (670nm) filter was inserted into the excitation arm to filter out the unwanted scattering laser light from the fiber. The collected light was passing through a long-pass (690nm) filter. A spectrometer with a CCD (charge-coupled device) cooled by liquid nitrogen was used to measure the PL. The spectrum was defined by 1500mm grating with the spectral resolution~70 µeV. Photon autocorrelation measurements were performed after collecting the signal into a multimode fiber beam splitter, where the counts were recorded by two single photon avalanche photodiodes. The correlation function for the pulse experiments was described in the following formula:

$$g^2(t) = y0 + \sum_{n=1}^{n=7} A_n * (\exp(\frac{t-t_n}{\tau_a})\Theta(-t) + \exp(-\frac{t-t_n}{\tau_b})\Theta(t)) \qquad (1)$$

where $\Theta(t)$ is the step function, $y0$ is the offset, $A_n$ is the peak height of the $n$th peak, $t_n$ is the peak position of the $n$th peak. $\tau_a = \tau_b$ in the exciton (X) autocorrelation and the biexciton (XX) autocorrelation measurement, which are related to characteristic radiation time of X and XX.

The CW cross correlation data in the inset of Fig.3c was fitted with multiexcitonic model similar to that in semiconductor QDs[34]:

$$g^2(t) = g^2(t, neg)\Theta(-t) + g^2(t, pos)\Theta(t) \qquad (2)$$

where $g^2(t,c) = 1 - A_{1c} * \exp(-|\frac{t}{t_1}|) - A_{2c} * \exp(-|\frac{t}{t_2}|)$. $\qquad (3)$

For $g^2(t<0)$, we will see the anti-bunching while the bunching for $g^2(t>0)$.

In order to account for the finite timing resolution of our setup, we fitted the pulse correlation data in Fig.3 with the correlation function convolved with a Gaussian distribution function:

$$f_{Det}(t) = \frac{1}{\sigma\sqrt{\pi}} e^{-\frac{1}{2}\left(\frac{t-t_0}{\sigma}\right)^2}, \quad (4)$$

where the $2\sqrt{2\ln 2}\sigma$ represents the 350ps APD time resolution. So the real fitting function would be:

$$g^2_{real}(\tau) = (g^2(t) * f_{Det})(\tau) = \int_{-\infty}^{\infty} g^2(t) f_{Det}(\tau-t) dt \quad (5)$$

Photon cross-correlation measurements were carried out after passing the signal into a 50:50 beam splitter. One beam was filtered out at P2 by a pair of 1nm bandwidth filters centered at 720nm and the other beam was leading to the monochromator with the filtered spectrum at P1. The pulsed $g^2(0)$ value was calculated from the fitted area in the zero-time delay peak divided by the average of the adjacent six peaks.

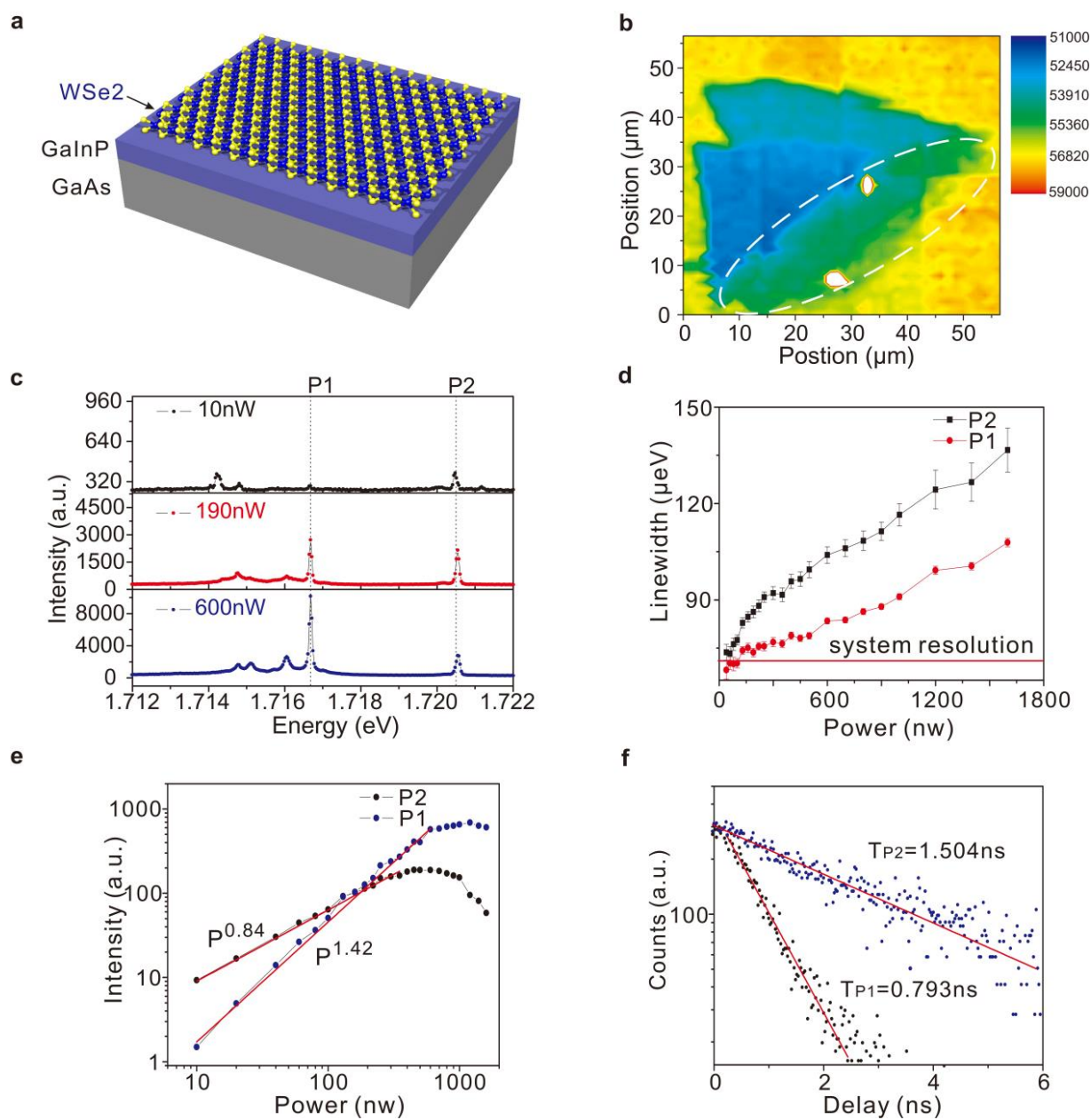

**Figure 1 | Narrow spectral lines in monolayer WSe$_2$. a**, Schematic illustration of the device with the deposited WSe$_2$ monolayer. **b**, Contour plot of Photoluminescence intensity within the energy range between 1.525eV and 1.734eV, over 56μm×56μm. The dashed white line marks the potential monolayer area. **c**, Photoluminescence spectrum of the localized emitter in the WSe$_2$ monolayer at 4.5K, showing different emission behaviors with increasing laser power dominated by peaks at 1.7167eV (P1) and 1.7206eV (P2). **d**, Extracted linewidth of P1 and P2, plotted as a function of excitation power. The spectrum for low excitation power shows resolution limited linewidth for both P1 and P2. **e,** The integrated counts of the photon emission from P1 and P2 shows super-linear and sub-linear behavior with increasing laser power in Log-Log plot. The red line is the power law fitting $I \propto P^x$, with the extracted x=0.84 for P2 and x=1.42 for P1. **f,** Time resolved photoluminescence of P1 and P2 show the

single exponential decay with the time constant of 0.793±0.017ns for P1 and 1.504±0.028ns for P2.

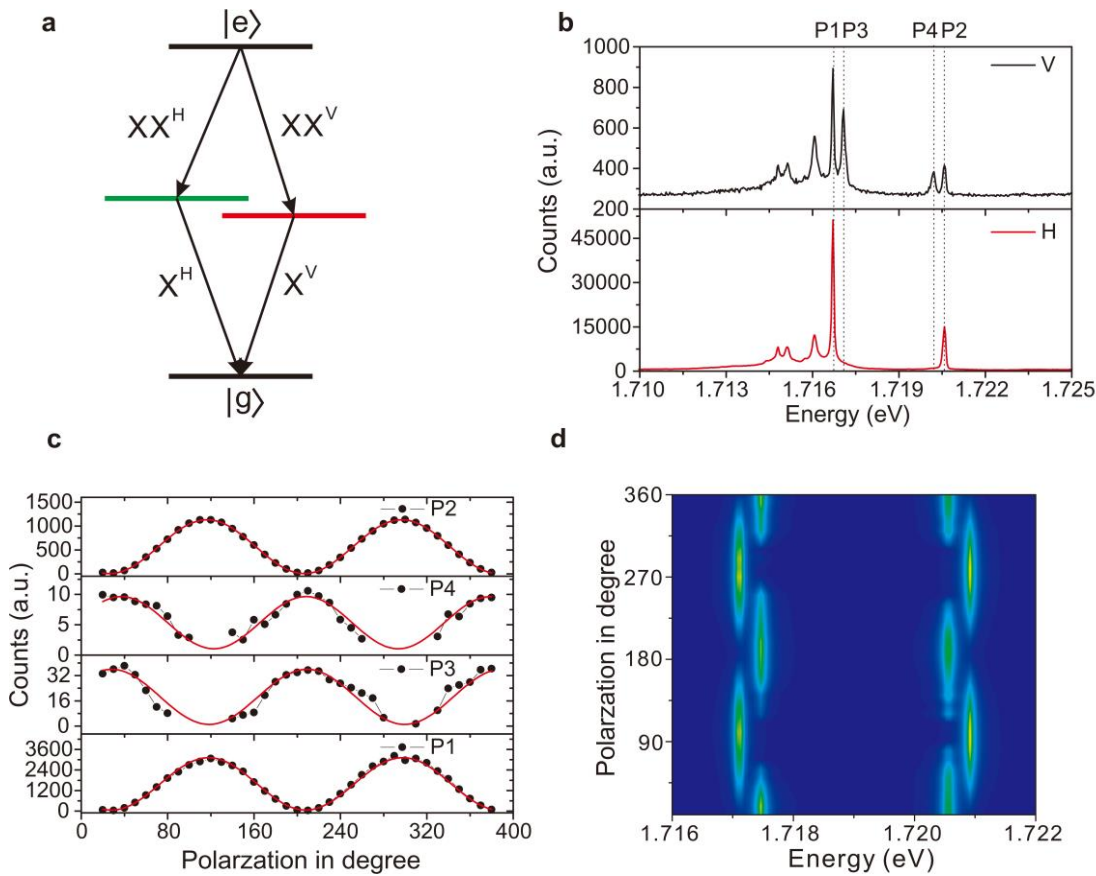

**Figure 2 | Polarization-resolved photoluminescence. a**, Schematic representation of biexcitonic emission cascade. The fine-structure splitting is expected for the electron-hole exchange interaction in presence of in-plane anisotropy. **b**, Polarization resolved spectrum at linear polarization H, V. Two pairs of spectral doublets are observed at 1.7167 (P1)-1.7171eV (P3) and 1.7202 (P4)-1.7206eV (P2). Four peaks are indicated by the dashed lines. **c,** The integrated counts of the photon emission from P1, P2, P3, P4 as a function of the polarization detection angle. The red lines are the sinusoidal fits, showing 2 pairs of cross linear-polarized doublets. **d,** Contour representation of the four peaks, after normalizing to the maximum peak intensity, yielding a fine structure splitting ~0.4meV.

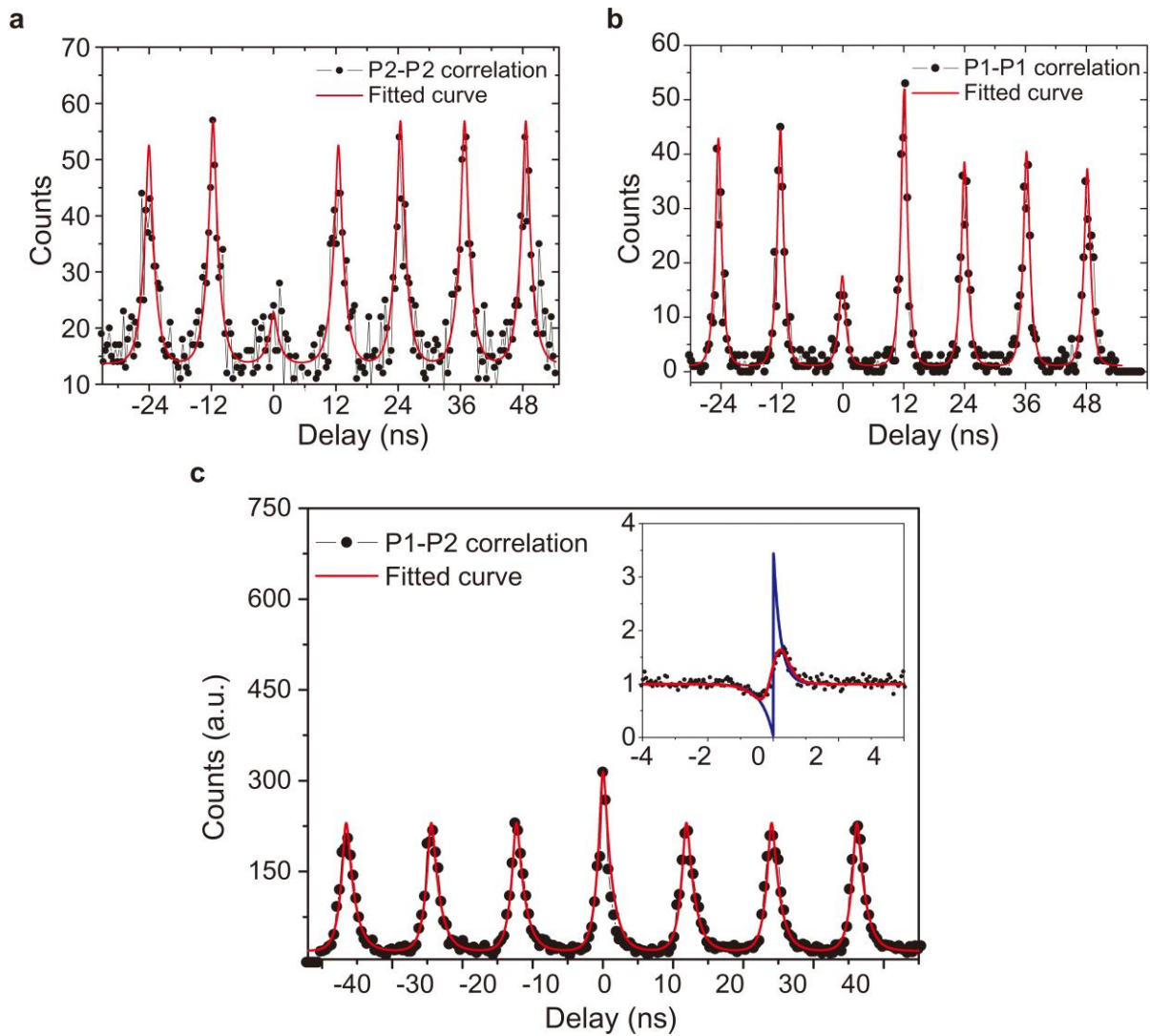

**Figure 3 | Photon correlation measurments.** **a,b** Second-order autocorrelation of measurement of the P2 and P1 under 13.6nW pulsed excitation at 475nm with the repetition rate of 82MHz and the pulse length of 3ps. We could extract $g^2(0)=0.397\pm0.06$ for P1 and $g^2(0)=0.286\pm0.04$ for P2. **c,** Cross-correlation measurement between P1 and P2 under pulsed excitation. The obtained $g^2(0)=1.404\pm0.038$, which obviously shows the bunching effect of the emission. The red lines are fitted with two exponential decay convolved with the response function. The inset is the cross correlation measurement of P1 and P2 under a 70nw CW laser excitation at 532nm. The red line in the inset is a fit with multiexcitonic model convolved with the response function. The blue line is the deconvoluted curve, which shows $g^2(0)=0.038\pm0.004$ for negative delay and $g^2(0)=3.434\pm0.129$ for positive delay. More details about fitting could see Methods.